\documentstyle[preprint,prd,floats,aps]{revtex}
\begin{document}
\tighten
\preprint{SU-GP-97/10-2}

\def\T{{\cal T}}

\input epsf.tex

\title{On the consistency of the constraint algebra in spin network quantum 
gravity}

\author{Rodolfo Gambini\footnote{Associate member of ICTP.}
\\{\em Instituto de F\'{\i}sica, Facultad de
Ciencias, Tristan Narvaja 1674, Montevideo, Uruguay}}

\author{
Jerzy Lewandowski\footnote{On leave from Instytut Fizyki Teoretycznej,
Uniwersytet Warszawski, ul. Ho\.za 69, 00-681 Warszawa, Poland}\\
{\em 
Max-Planck-Institut f\"ur Gravitationsphysik
Schlaatzweg 1
D-14473 Potsdam, Germany}}

\author{Donald Marolf
\\{\em Physics Department, Syracuse University, Syracuse, NY 13244-1130}}

\author{Jorge Pullin\\  {\em Center for Gravitational Physics and Geometry,
Department of Physics,\\ 104 Davey Lab, The Pennsylvania State 
University, University Park, PA 16802}}

\maketitle
\vspace{-7.5cm} 
\begin{flushright}
\baselineskip=15pt
CGPG-97-10/1\\
gr-qc/9710018\\
\end{flushright}
\vspace{7.5cm}

\begin{abstract}
We point out several features of the quantum Hamiltonian
constraints recently introduced by Thiemann for Euclidean gravity.
In particular we discuss the issue of the constraint algebra and of the 
quantum realization of the object $q^{ab}V_b$, which is classically the
Poisson Bracket of two Hamiltonians.
\end{abstract}

\section{Introduction}

In a remarkable series of papers \cite{QSD1,QSD2,QSD3,QSD} by
Thiemann,  the loop
approach to the quantization of general relativity reached a new level.
Using ideas related to those of Rovelli and Smolin \cite{RoSm},
Thiemann proposed a definition of the quantum Hamiltonian
(Wheeler-DeWitt) constraint of Einstein-Hilbert gravity which is a
densely defined operator on a certain Hilbert space and which is
(in a certain sense \cite{QSD1}) anomaly
free on diffeomorphism invariant states.  The fact that the
proposed constraints imply the existence of a self-consistent, well
defined theory is very impressive.  However, it is still not clear
whether the resulting theory is connected with the physics of 
gravity. This question has been raised in \cite{Lee,LeMa}, and this
paper intends to probe it further.

We will address two issues. 
The first concerns the commutator of two Hamiltonian constraints.
This topic
has been partially considered by Thiemann \cite{QSD3} and
in the preceding paper \cite{LeMa} by Lewandowski and Marolf.
Thiemann argued \cite{QSD1}
that the commutator of two Hamiltonians vanishes on diffeomorphism
invariant states by commuting two `regulated' constraints and showing
that, for sufficiently small regulators, the result 
annihilates diffeomorphism invariant states.
He also considered \cite{QSD3} the function appearing on the
right hand side of the classical Poisson bracket of two Hamiltonians:
\begin{equation}
\label{algebra}
\{H(N),H(M)\} = \int d^3x (N\partial_b M -M\partial_b  N) \tilde{E}^c_i 
F_{cb}^iq^{ab}
\end{equation}
where  the
Hamiltonian constraint is 
\begin{equation}
\label{EH}
H(N) = \int d^3x N \epsilon_{ijk}
E^a_i E^b_j F_{ab}^k.
\end{equation}
Our notation is that $E^a_i$ is the spatial
triad while $\tilde{E}^a_i$ is the densitized triad, $F_{ab}$
is the field strength of the pull back of the (Euclidean)
self-dual curvature to a spatial hypersurface, and $\{,\}$
denotes the Poisson bracket. Thiemann was able
to show that a quantum
version of the
right hand side of (\ref{algebra}) annihilates
diffeomorphism invariant states.  This is the result one would expect from 
classical reasoning as
\begin{equation}
V(N) := \int N^b \tilde E_i^c F^i_{cb} =: \int N^b V_b
\end{equation}
is the classical function that generates diffeomorphisms.  Thus, modulo
factor ordering issues and anomalies, the quantum commutator
corresponding to (\ref{algebra}) should
annihilate diffeomorphism invariant states.

In summary, Thiemann showed that the algebra is
consistent at this level: his quantum versions of both the left and right
 hand sides of (\ref{algebra}) vanish 
on diffeomorphism invariant states.  On the other hand,
Lewandowski and
Marolf \cite{LeMa} explored the commutator of two of Thiemann's
quantum Hamiltonians on a larger space 
(called ${\cal T_*'}$) of states that 
are not necessarily
diffeomorphism invariant. It was found that the 
commutator of the Hamiltonian constraints continues to vanish, even
on this larger space.
Such a result was also shown to hold for a large
class of extensions \cite{QSD2,DPR} of the original proposal
of \cite{QSD1}. This suggested that an inconsistency might
appear, since  
the classical phase space function on the right hand side of
(\ref{algebra}) vanishes
only when the diffeomorphism constraint is satisfied.
Following Thiemann \cite{QSD3}, let us denote this function by
${\cal O}(N,M)$:
\begin{equation}
\label{O}
{\cal O}(N,M) = \int d^3 x (N \partial_a M - M \partial_a N )  V_b q^{ab}.
\end{equation}   
We will discuss quantum versions
of ${\cal O}(N,M)$ in this paper.
In particular we will show that the original quantization of this function
proposed by Thiemann also yields the zero operator on the
larger space ${\cal T}'_*$.  As a result, 
one could consider the calculation consistent.  However, 
the price for this consistency appears to be ``setting $q^{ab} =0$"
as an operator
on this space of states (more on this later).
This is so in spite of the fact that
${\cal T}_*'$ contains both the
diffeomorphism invariant states of \cite{QSD1,QSD2} and, in particular, the 
`physical' states that solve the Hamiltonian constraints.
We will also show that there exist regularization
ambiguities that allow us to propose quantum versions of $q^{ab}$
and ${\cal O}(N,M)$ that are nonzero. However,
the quantum commutator still vanishes, so the algebra
is anomalous with this new proposal.  This is discussed in section II.

The second issue 
concerns the computation of the commutator
presented in the companion paper \cite{LeMa} in
light of the previous point: would it be possible to redefine the constraints
themselves in such a way as to get a non-zero result?
And if so, would the result coincide with some nontrivial
quantum version of ${\cal O}(N,M)$? From
addressing these questions,  section III will derive a rather general property
that is necessary (but not necessarily sufficient) for 
operators of this type to
reproduce the classical hypersurface deformation algebra. 

The following treatment works
within the loop approach to (Euclidean) quantum gravity and
uses the now standard technology of spin networks and
linear functionals thereon.  We refer the reader to the 
companion paper \cite{LeMa} for a complete description of the
context and conventions employed here.  Some of the founding works in
the field are \cite{JS,NP}.  For a thorough description of
this approach, see \cite{AA,ALMMT}.  For a description of spin 
networks, see \cite{Baez} and \cite{SN}.

\section{Commutator of two Hamiltonians}

The classical Poisson bracket of two Hamiltonian constraints, equation
(\ref{algebra}), indicates that the commutator is proportional to a
diffeomorphism generator. 
However,  the proportionality is through an operator-valued
quantity involving the metric. If one has a Hamiltonian constraint
defined on diffeomorphism invariant states, its commutator should
therefore vanish on such states if the factor ordering is such that
the diffeomorphism acts on the state first. Thiemann showed 
from the outset \cite{QSD1} that the commutator of his (properly regulated)
constraints annihilates
such states. In fact, it is not
difficult to see. The action of Thiemann's constraint on a spin network
state $|\Gamma \rangle$ is 
to generate a number of terms, each of which is a single
spin network resembling $|\Gamma \rangle$ but with an extra edge added
near an appropriate intersection 
of the original spin network, and with the
spins and contractor changed at that intersection.
Two successive actions of
such an operator 
add a total of two edges in each term and the order in which the operators are
applied affects only the position occupied by the new lines.  In particular, 
it does not effect the overall topology (or diffeomorphism class) 
of the new graph or the spins or contractors.  Thus, 
the result is independent of which constraint is applied first 
if one is able to ``slide the added
lines'' back and forth.  Sliding lines in this way is allowed when 
we compute $\langle \psi | \hat{H}(N) |\Gamma \rangle$
for  diffeomorphism invariant states $\langle \psi|$ in an appropriate
dual space.  As a result, the {\it dual} action of the
commutator on `bra' states annihilates all such diffeomorphism invariant
states $\langle \psi |$.  In this sense then, the constraints presented 
have a consistent algebra on these states.

The previous paper \cite{LeMa} considered a variation of this construction. We
describe it only briefly here and refer the reader to the original work
for further details.   The basic idea was to consider 
an enlarged space of states, whose members need not be invariant
under diffeomorphisms.  We now consider a large subspace 
of this space 
which will be sufficient for our purposes.  This space
is constructed as follows.

Suppose that we work on a three manifold $\Sigma$ and consider a spin
network $\Gamma = (\gamma, j, c)$ which we think of as embedded in $\Sigma$.
In this notation, $\gamma$ is graph, the label $j$ assigns a representation
of $SU(2)$ to each edge of the graph, and the contractors $c$ assign a 
certain type of operator to each vertex in $\gamma$.
Suppose that the associated graph $\gamma$ has no symmetries in
the sense of $\cite{LeMa}$.  That is, the only diffeomorphisms of
$\Sigma$ that map $\gamma$ to itself also map each edge of
$\gamma$ to itself without changing the edge's orientation.  The
reader may consult \cite{LeMa} for the more general case.  

We will not just leave this
spin network at some fixed location in $\Sigma$, but
will transport it to other locations in the manifold.  In fact, 
given any map $\sigma$ which assigns a point of $\Sigma$ to each
vertex  of $\gamma$,
we introduce a linear functional $\langle \Gamma_\sigma|$
on the space (${\cal T}$) spanned by spin networks.  To define
this functional, suppose that we have another spin network
$\Gamma' = (\gamma',j',c')$ for which $\gamma'$ is related to $\gamma$
by a smooth diffeomorphism $\varphi$,  $\gamma' = \varphi(\gamma)$, for
some smooth diffeomorphism $\varphi$ that takes
the vertices of $\gamma$ to just those points assigned by $\sigma$.
Then we set $\langle \Gamma_\sigma | \Gamma' \rangle := \langle \Gamma
| {\cal D}_{\varphi^{-1}}|\Gamma' \rangle$, where ${\cal D}_\phi$ 
denotes the unitary action of a diffeomorphism $\phi$ on spin
network states and the matrix elements 
$\langle \Gamma |{\cal D}_{\varphi^{-1}}
|\Gamma' \rangle$ are computed using the inner product of 
$L^2(\overline{A/G},
d\mu_0)$ for the Ashtekar-Lewandowski measure $d\mu_0$ of \cite{ALM}.
For any other $\Gamma'$ (over the wrong sort of graph or in the
wrong location) we set $ \langle \Gamma_\sigma | \Gamma' \rangle : = 0$.

The states of interest are those obtained by `smearing out' these
states with some smooth function $f$ on the space $\Sigma^{V(\gamma)}$ of
all maps from the vertices $V(\gamma)$ of $\gamma$ to our three
manifold $\Sigma$.  That is, we are interested in states $ \langle
\Gamma, f|$
of the form
\begin{equation}
\label{smoothdef}
\langle \Gamma, f| = \sum_{\sigma \in \Sigma^{V(\Gamma)}}
\langle \Gamma_\sigma|
f(\sigma).
\end{equation}
While this sum involves an (uncountable) infinity of terms, its action
on spin network states $|\Gamma' \rangle$ is well-defined since only 
one term (the one in which $\sigma$ maps the vertices of $\Gamma$ to
the vertices of $\Gamma'$ in the proper way) can be nonzero.  So, 
this sum basically says that $\langle \Gamma, f|$ acts on $|\Gamma' \rangle$
by moving $\Gamma$ `on top of' $\Gamma'$ and then taking the inner
product of $|\Gamma\rangle $ and $|\Gamma' \rangle$ in $L^2(\overline{A/G},
d\mu_0)$, weighted by the function $f$.  In particular, $\langle \Gamma, f|$
is diffeomorphism invariant whenever $f$ is a constant function.  If
the vertices of $\Gamma$ are numbered from $1$ to $k$, then  
we may rewrite the sum (\ref{smoothdef}) as
\begin{equation}
\langle \Gamma, f | = \sum_{x_1,...,x_k \in \Sigma} \langle \Gamma_{x_1,...x_k}
| f(x_1,...,x_k)
\end{equation}
with the obvious correspondence between $\sigma$ and $(x_1,...,x_k)$.
Note that $f$ is required to be a smooth function on the entire
space $\Sigma^k$.  The space ${\cal T}'_*$ of \cite{LeMa}
consists of arbitrary superpositions 
of the $\langle \Gamma, f|$, together with the corresponding 
extensions to graphs with symmetries.  Again, such superpositions
may include an uncountable infinity of terms, so long as all but a finite
number annihilate any give spin network.  The constraints of \cite{QSD1}
are well-defined on this space, as are the variations of \cite{QSD2,DPR,Lee}.
Such constraints also map ${\cal T}'_*$ to itself, so that commutators
can be calculated directly.  Note that
the generator of diffeomorphisms is well-defined and
nonzero, and acts just by Lie dragging the function $f$.

At the intuitive level (and in the typical case), 
it is not difficult to see that Thiemann's constraint is well defined
on this space of states and that it has a vanishing commutator with
itself.  For simplicity, suppose that we evaluate the
`matrix elements' $\langle \Gamma, f| [H(N),H(M)]|\Gamma' \rangle$
of the commutator between
the state $\langle \Gamma, f|$, and the spin network $|\Gamma' \rangle$, 
where $\Gamma$ has only one vertex $x$.
The idea is
that the successive action of two Hamiltonian constraints will add 
new lines at the intersection of the spin network, but the smearing
functions $N$ and $M$ get evaluated at exactly the same point $x$, (the
original vertex). Thus, one
gets a single term with a pre-factor of the form $N(x) M(x)-M(x) N(x)$ 
and the commutator vanishes. When $\Gamma'$ has multiple vertices, these
vertices do not interact with each other and the commutator continues to
vanish.   This is true \cite{LeMa} even if one uses 
the version of the volume operator due to
Rovelli and Smolin \cite{RoSmvo}
in the definition
of the Thiemann Hamiltonian. In that case, the second Hamiltonian
acts not only at the original intersection 
but also at the newly added vertices (the Ashtekar--Lewandowski
volume does not act on those vertices because they are planar, but the
Rovelli--Smolin volume does not automatically vanish on planar
vertices)\footnote{It has been pointed out by Roberto De Pietri 
\cite{DePico} that, because Thiemann proceeds by triangulating the
manifold $\Sigma$ with nondegenerate tetrahedrons, the planar
vertices are explicitly removed from consideration.  Strictly
speaking, this represents a breakdown of the arguments of \cite{QSD1} when used
with the Rovelli-Smolin volume operator.  However, it is
straightforward to generalize the argument so that the
Rovelli-Smolin volume can be used and so that the Hamiltonian acts
non-trivially at planar vertices.}.
In this case, one gets a contribution with a prefactor
$N(x) M(y)-M(x) N(y)$ where $x$ is the location of the first
intersection and $y$ the one of the second. However, 
one removes
the regularization by taking
$x\rightarrow y$ and the commutator vanishes for smooth $N$ and $M$.

Clearly, this should be 
a source of worry. Why does the commutator
vanish on non-diffeomorphism invariant states? However, it could be that
the states considered are not general enough and therefore  that
the commutator just happens to vanish on these states.
To analyze this question, one should look at
a quantization of ${\cal O}(N,M)$. It is rather difficult to realize a quantum version of
the right hand side
in the spin network representation, due to the presence
of the inverse metric.
Nevertheless, it can be done, as discussed in detail by Thiemann
\cite{QSD3}. The idea is to break up
the metric into two un-densitized contravariant triads $q^{ab} =e^a_i
e^b_i$, which in turn can be related to co-triads via,
\begin{equation}
\label{e1}
e^a_i = \epsilon^{abc} \epsilon_{ijk} {e^j_b e^k_c\over \sqrt{det(q)}}
\end{equation}
and finally the co-triads can be expressed in terms of a commutator of
a connection and a volume,
\begin{equation}
\label{e2}
{1 \over 2} {\rm sgn}(det(e)) e^i_a = \{ A_a^i,V\}
\end{equation}
and $V=\int d^3x
\sqrt{\epsilon^{abc}\epsilon_{ijk} 
\tilde{E}^a_i \tilde{E}^b_j\tilde{E}^c_k}$ is the volume of the three
manifold.   

We refer the reader to Thiemann's paper \cite{QSD3}
for the details, but the main
point is that the resulting quantum operator for the right-hand side
of the constraint can be written as a finite well defined operator
times a prefactor in which
the smearing functions $N$ and $M$ are
evaluated at the intersection point $v$ where
the Hamiltonian acts and at the end of an outgoing link $\Delta$ from $v$;
the exact form is $M(v) N(\Delta) - M(\Delta) N(v)$ in the notation of
\cite{QSD3}. In the limit
in which the regularization is removed, the length of the segment
$\Delta$ shrinks to zero and the prefactor vanishes for continuous $N,M$.
We have therefore
shown that the quantization of ${\cal O}(N,M)$ proposed by Thiemann
coincides with the commutator derived by Lewandowski and Marolf 
on the particular space of states ${\cal T}'_*$ defined above: both
operators vanish.
Therefore, there is no anomaly or inconsistency at this level.
However, an inspection of eq. (3.7) from  paper \cite{QSD3} shows that
the $M(v)N(\Delta)-M(\Delta)N(v)$ term in fact arises in the 
regularization  from the factor $(MN_{,_a}-NM_{,a})q^{ab}$. 
Since the vector constraint factor $V_b$ may be treated as simply
the infinitesimal diffeomorphism generator, the procedure of \cite{QSD3}
is tantamount to a regularization of $q^{ab}$ which sets $q^{ab}=0$
on ${\cal T}_*'$.
Note that, while $q^{ab}$ is not diffeomorphism invariant, this does not mean 
that it cannot be defined as an operator on some space of states (such
at $\T$) whose members are not diffeomorphism invariant. 

It should also be pointed out that Thiemann \cite{QSD5} has proposed
a general scheme for regularizing a large class of operators that would
include $\sqrt{\det q} q^{ab}$ but not $q^{ab}$ itself.  This 
procedure appears to give a nonzero $\sqrt{\det q} q^{ab}$ operator.
However, the procedure of \cite{QSD5} differs from the procedure used
to regularize $O(M,N)$ in \cite{QSD3}.  In particular, if the procedure
of \cite{QSD5} were applied to $O(M,N)$ it would not yield an
intact factor of the vector constraint $V_b$ on the left hand side
of $O(M,N)$.  As a result, the relationship between the operators resulting
from these different schemes is unclear.  Further study of this 
issue would be worthwhile.

Let us therefore attempt to understand just why 
the aforementioned quantization of $q^{ab}$ yields the zero operator.
The root of the issue seems to 
lie in the fact that this quantization 
involves the prefactor $M(v) N(\Delta) - M(\Delta) N(v)$, 
involving the smearing functions evaluated at ``close''
points $v$, $\Delta$,
but does not produce a corresponding denominator that goes to zero when the
regulators are removed and $\Delta \rightarrow v$.
Therefore one always fails to reconstruct the
derivatives that appear in the classical expression (\ref{algebra})
and one gets a
vanishing result.

This feature can in turn be traced to
the quantization of expressions of the form:
\begin{equation}
\label{trouble}
\{A^i_a(x),\sqrt{V(x,\epsilon)}\} = {\{A^i_a(x),V(x,\epsilon)\} \over 2
\sqrt{V(x,\epsilon)}}
\end{equation}
\noindent that were introduced through equations (\ref{e1}) and 
(\ref{e2}).  Here,  
\begin{equation}
V(x,\epsilon) = \int d^3x
\sqrt{\epsilon^{abc}\epsilon_{ijk} 
\tilde{E}^a_i \tilde{E}^b_j\tilde{E}^c_k}\chi(\epsilon,x)
\end{equation}
is the metric
volume of a box with center $x$ and size $\epsilon$
and $\chi$ is the characteristic function of  this box.
Factors of this type arise in the constructions of
\cite{QSD3}.   Classically,  expression (\ref{trouble}) diverges
in the limit $\epsilon\rightarrow 0$: the denominator
vanishes since it is the volume of a box of size $\epsilon$, and the
numerator is independent of $\epsilon$ because the Poisson bracket
is local.  This divergence is, of course, canceled by other factors of
$\epsilon$  so that Thiemann's classical expression for ${\cal O}(N,M)$ is
finite.   However, Thiemann's quantization \cite{QSD3} 
of (\ref{trouble}) has a {\it finite} action at intersections of spin
networks.  The point is that he quantizes this expression
by promoting the Poisson Bracket on the left to a commutator (and
replacing the connection with a holonomy).  Since his operator
$\hat{V}(x,\epsilon)$ is finite 
as $\epsilon \rightarrow 0$, the commutator is finite as well.
As a result, the other factors of
$\epsilon$ cause his quantum version of ${\cal O}(N,M)$ and the 
corresponding version of $q^{ab}$ to vanish.
This suggests that the quantization is not faithfully
recovering properties of the classical operator.

It turns out that, within the general quantization scheme described
in \cite{QSD1,QSD2,QSD3} there is in fact enough freedom to alter
this result.
Consider again the classical functions (\ref{trouble}).
Since the Poisson bracket $\{A_a,V(x,\epsilon)\}$ is
really $\epsilon$ independent, one could replace $V(x,\epsilon)$ 
with some $V(x,\delta)$ in the Poisson bracket $\{A^i_a(x), V(x,\epsilon)\}$, 
while keeping the same regulator $\epsilon$.
Furthermore, 
for small $\epsilon$, $V(x,\epsilon)$ scales as $\epsilon^3$ for
smooth fields.   If one now 
reworks the above argument keeping this in mind, one may rewrite 
equation (\ref{trouble}) as
\begin{equation}
\widehat{{\{A,V(x,\delta)\} \over \sqrt{V(x,\epsilon) }}} = 
2  \left({\delta \over
\epsilon}\right)^{3\over 2} \{\hat{A},\hat{V}^{1/2}(x,\delta)\}.  
\end{equation}

The point is that, in the proposed quantization, $\lim_{\delta
\rightarrow 0}\hat{V}(x,\delta)$ yields a well-defined finite
operator, which is therefore independent of $\delta$.
It is then evident\footnote{One needs to choose dimensionless
coordinates; if not one needs to add extra dimensional factors.} that,
by choosing $\delta=\epsilon^n$, one can generate extra
powers of $\epsilon$ in the the construction of the quantum ${\cal O}(N,M)$. 
In particular, because four factors of the form
(\ref{trouble}) are used to write $q^{qb}$, the choice of $n=5/6$ 
provides the single inverse
power of epsilon needed to convert the difference $M(v) N(\Delta)
- N(v) M(\Delta)$ into
the structure that appears in (\ref{algebra}).
With such a choice one gets
a non-zero operator $q^{ab}$ which is well-defined on $\T'_*$, 
although it is clearly not unique.

One could imagine using similar tricks to rescale the quantum
constraints so that their
commutator is non-vanishing. We will consider the
resulting expressions in the next section. However these manipulations
are only formal in character.  In order to yield 
a non-vanishing result, the commutator must be rescaled by $1/\epsilon$.
This, however,  requires
each Hamiltonian to be rescaled by $1/\sqrt{\epsilon}$.  Since the
original constraints $\hat{H}(N)$ were nonzero and well-defined, the
new constraints are roughly $\hat{H}(N)/\sqrt{\epsilon}$ and are
divergent as $\epsilon \rightarrow 0$.
Whether or not this
can be dealt with through some sort of renormalization 
is unclear.

\section{Computing the commutator}

Ignoring for the moment the issue raised in the last paragraph, one
can ask what would be obtained if we would compute the commutator of
two regulated Hamiltonians and rescale it by one power of $1/\epsilon$
before taking the limit $\epsilon \rightarrow 0$.  We will not worry
here about the details of how this is done.  Although Thiemann does
not explicitly use expressions of the form (\ref{trouble}) in writing
his constraints, they may be introduced through suitable
manipulations.  Modulo perhaps extra powers of the regulator
$\epsilon$, the result is again an `RST-like'
(Rovelli--Smolin--Thiemann-like) operator in the sense of \cite{LeMa}.
As a result, if there were no extra $\epsilon^{-1}$'s, the commutator
of constraints would still be zero \cite{LeMa}.

If the $\epsilon \rightarrow 0$
limit of the rescaled commutator
is properly taken, one would expect it not to vanish.  The point
of such a computation is that one might think that extra powers of
$\epsilon$ could be associated with some sort of renormalization.
Thus, without studying the details, we address the question of whether
any such renormalization is likely to yield the classical algebra.
The importance of the calculation lies in general properties that one may 
recognize as contributing to reproducing ---or failing to 
reproduce---
the result expected from the classical theory. 
Exploring the theory in this way before removing regulators is 
reminiscent of manipulations performed in a lattice context \cite{lat}.

Instead of beginning the calculation immediately, let us first
discuss the {\it desired} result.  We would hope to find some
analogue of the classical function ${\cal O}(N,M)$.
Now, if there are no anomalies, one would
expect the quantum commutator  also to resemble (\ref{O}) and, what is more, 
to have the factor ordering implied by (\ref{O}).  This is because
the constraints of \cite{QSD1} are to be applied on `bra'
states $\langle \Gamma, f|$ so that, in order for the algebra to
close properly, the vector constraint $V_b$ should appear on the left.

Recall now that $V_b$ are the diffeomorphism generators and that 
the action of diffeomorphism generators are well defined on
$\T'_*$.  Given a vector field $N^a$, the action of an infinitesimal
diffeomorphism along $N$ on the state $\langle \Gamma, f|$ 
yields just the state $\langle \Gamma, {\cal L}_Nf|$ with the
same spin network $\Gamma$ smeared against the Lie derivative ${\cal L}_Nf$
of $f$ along $N$.  We note that this state takes the form
\begin{equation}
\langle \Gamma, {\cal L}_Nf| = \sum_{\sigma \in \Sigma^{V(\Gamma)}}
\left( \langle \Gamma_\sigma| \sum_{x \in V(\Gamma) }  N^a(\sigma(x))
{{\partial f} \over {\partial (\sigma(x))^a}} (\sigma) \right).
\end{equation}
As a result, one would expect
matrix elements of the quantum commutator to be of the form
\begin{eqnarray}
\label{expect}
&&\langle \Gamma, f | [ \hat{H}(N), \hat{H}(M) ] | \Gamma' \rangle
=  \cr &&
\sum_{\sigma \in \Sigma^{V(\Gamma)}} 
\left( \sum_{x \in V(\Gamma) } (N (\sigma(x)) \partial_a M (\sigma(x))
 - M(\sigma(x)) \partial_a(\sigma(x)) N) 
\langle \Gamma_\sigma| \widehat{ q^{ab}} (x) | \Gamma'
\rangle  
{{\partial f} \over {\partial (\sigma(x))^b}}(\sigma) \right)
\end{eqnarray}
if in fact the action of the operator $\widehat{q^{ab}}(x)$ commutes 
with the sum over
$\sigma$.   Here, $\widehat{q^{ab}}(x)$ is some quantization of $q^{ab}(x)$.
The details of this operator, and therefore of (\ref{expect}), will
depend on the form of this operator.
Still, one expects each
term to have a prefactor $(N \partial_a M - M \partial_a N)$ and to 
involve a derivative of $f$.  Also, for each vertex $x \in V(\Gamma)$,
one would expect to find a term involving ${{\partial f} \over { \partial 
\sigma(x)}}$. 

For simplicity we will  calculate the commutator only on a single
state $\langle \Gamma, f|$, where $\Gamma$ and a labeling of its
vertices is shown in figure 1.  
\begin{figure}
\centerline{\epsfbox{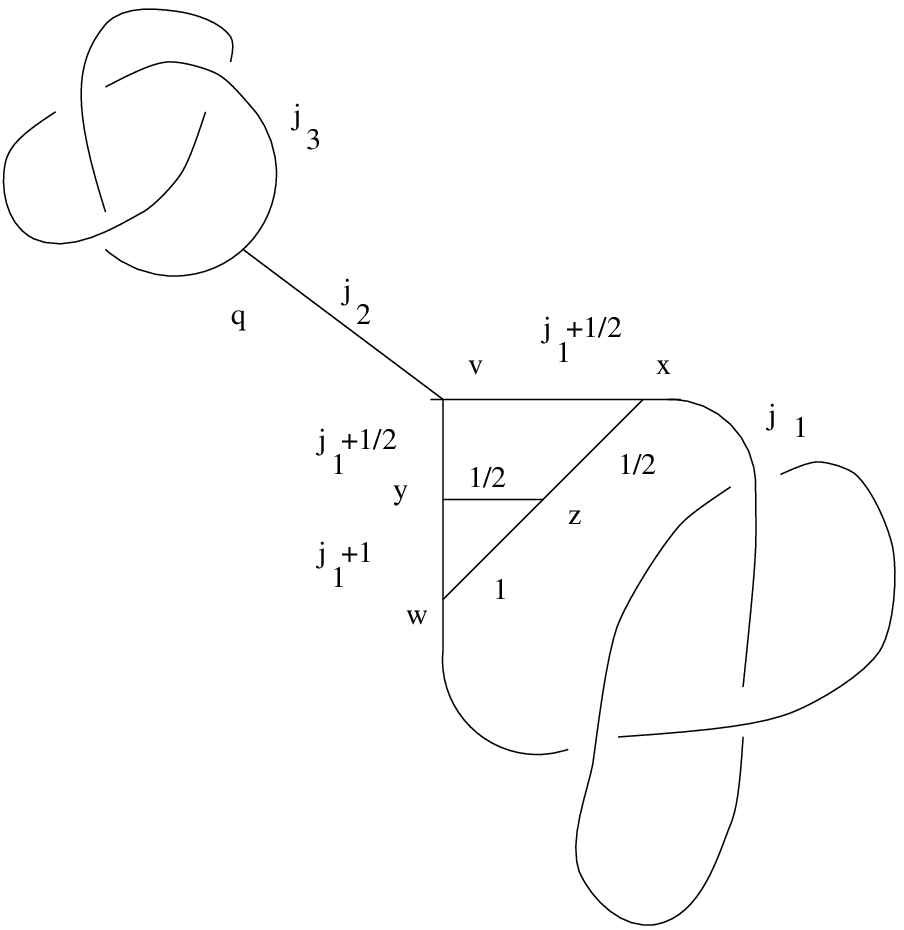}}
\vspace{2mm}
\caption{The type of spin network state on which we act with the commutator
of two Hamiltonians. Our notation is that $j_1,j_1+1/2, J_1+1,J_2,J_3,1,1/2$
are spins that label edges while $q,v,w,x,y,z$ are vertices of the 
graph.}
\end{figure}

Since $\Gamma$ has only trivalent vertices, we may fix the contractors
simply by requiring them to be projectors.  Also note that the
edge labeled by the spin $j_3$ forms a trefoil knot, as does
the loop formed by traveling from $x$ to $z$ to $w$ along the straight
edges and then closing the loop along the edge labeled by $j_1$.
Furthermore, these trefoil knots are of opposite chiralities, so that
neither the graph $\gamma$ underlying $\Gamma$ above
nor the graph obtained from it by removing
the straight edges connecting $x,z,y,w$ have nontrivial symmetries in the
sense of \cite{LeMa}.

As we do not know what to expect for the matrix elements $\langle \Gamma_\sigma
| \widehat{q^{ab}} | \Gamma' \rangle$ in (\ref{expect}),
our calculation of the commutator
will be only schematic.  We will not concern ourselves
with the details of spins and contractors, but merely note how
the lapse functions $N,M$ and the function $f$ are evaluated so that 
we can look for the derivatives which appear in (\ref{expect}).

To make the action of the Hamiltonian more interesting, we
will consider the case in which the volume operator is the
Rovelli--Smolin one. The calculation for the 
Ashtekar--Lewandowski volume is similar, but yields only a subset
of the terms we will obtain\footnote{However, a small change in
the the construction of \cite{QSD1} yields a constraint which acts
at planar vertices, even though it is built from the Ashtekar-Lewandowski
volume \cite{LeMa}.  The results for this new operator are then
schematically the same as for the Rovelli-Smolin case.}. Let us consider
a bra state of the form $\sum_{v,w,x,y,z,q} 
f(v,w,x,y,z,q) <\Gamma_{v,w,x,y,z,q}|$ where
$\Gamma$ is shown in figure 1. We now act with two Hamiltonians
$\hat{H}^\alpha(N)$ and $\hat{H}^\beta(M)$. By ``$\alpha$'' and ``$\beta$''
we mean two
possibly distinct regularizations of the Hamiltonian. The action of the
two Hamiltonians consists in fixing the vertices of the spin network
$w,x,y,z$ at points determined by the regularizations. 
The first constraint can act only at the vertex $w$ and removes
the spin $1/2$ edge between $y$ and $z$.  The second can only
act at the vertex $v$, and removes the edges between $x$ and $z$ and between
$z$ and $w$.  Since we act with regulated constraints, when the first
edge is removed, the corresponding arguments $y,z$ of $f$ are
evaluated at points $y^\beta,z^\beta$ (or $y^\alpha, z^\alpha$) that
depend on the regulator $\alpha$ ($\beta$) and of course implicitly
on the positions of the remaining vertices.  Similarly, when the second
edge is removed, the arguments $w,x$ are evaluated at some $w^\alpha,
x^\alpha$ ($w^\beta,x^\beta$). 

The end result is,
\begin{eqnarray}
\label{comm}
\sum_{v,w,x,y,z,q}&&
f(v,w,x,y,z,q) <\Gamma_{v,w,x,y,z,q}| (\hat{H}^\beta(N) \hat{H}^\alpha(M)-
\hat{H}^\alpha(M) \hat{H}^\beta(N)) \cr
&=&\ A(j_1,j_2) \sum_{v,q} \Big[ f(v,w^\alpha,x^\alpha,y^\beta,z^\beta,q)
N(w^\alpha) M(v)
\cr 
\ &-&\ f(v,w^\beta,x^\beta,y^\alpha,z^\alpha,q) M(w^\beta) N(v)
\Big] \langle \Gamma_{v,q}|
\end{eqnarray}
where $\langle \Gamma_{v,q}|$ is shown in figure 2 and $A(j_1,j_2)$
is some constant that depends on $j_1$ and $j_2$.

\begin{figure}
\centerline{\epsfbox{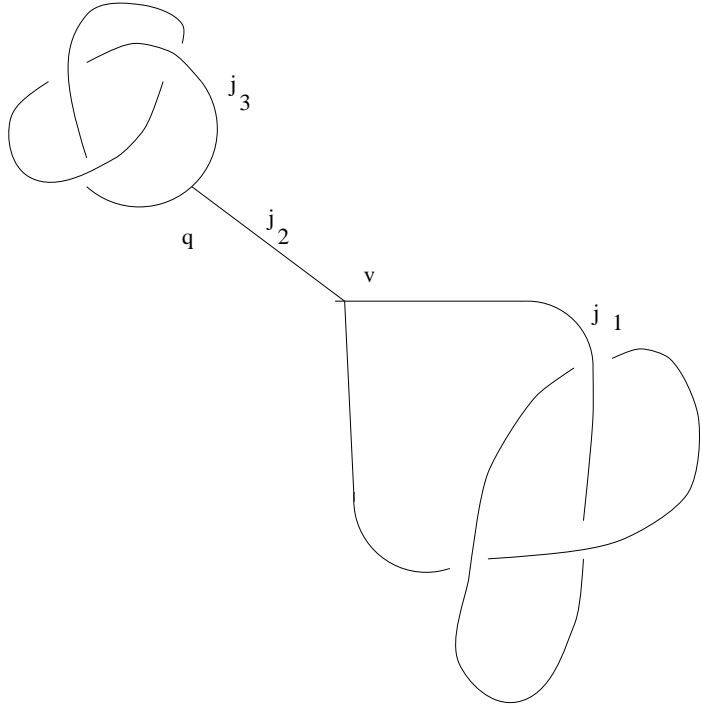}}
\vspace{2mm}
\caption{The type of state that results from the action of two Hamiltonians
on a state of the form shown in figure 1. As we see, the action of the 
Hamiltonians is to ``eliminate'' lines from the vertices of the graph.}
\end{figure}
If we now take
\begin{eqnarray}
|y^\beta -v|, |z^\beta-v| &\sim& \epsilon^{\alpha, \beta}, \cr
|w^\alpha -v|, |x^\alpha-v| &\sim& \epsilon^{\alpha}, \cr
|y^\alpha -v|, |z^\alpha-v| &\sim& \epsilon^{\beta, \alpha}, \cr
|w^\beta -v|, |x^\beta-v| &\sim& \epsilon^{\beta},
\end{eqnarray}
for small parameters $\epsilon^{\alpha,\beta}$, $\epsilon^\alpha$, 
$\epsilon^{\beta, \alpha}$, and $\epsilon^{\beta}$, the result
is schematically of the form
\begin{equation}
\label{smallsep}
\sum_v  \left[ (M \partial N 
\epsilon^\alpha -N \partial M \epsilon^\beta) f + M N \partial f 
(\epsilon^\alpha 
-\epsilon^\beta + \epsilon^{\beta,\alpha} - \epsilon^{\alpha,
\beta}) \right] \langle \Gamma_v|.
\end{equation}
In particular, because (\ref{smallsep})
involves evaluating $fNM$ at only two sets of arguments, it 
corresponds only to a set of terms having only one derivative each.
Thus, there are no terms like those found in (\ref{expect}), which 
contain products of $\partial f$ with $\partial N$ or $\partial M$
and so involve two derivatives of $f$, $N$, $M$.

If one had used the Ashtekar--Lewandowski volume, then in fact
$\hat{H}^\beta(N) \hat{H}^\alpha(N)$ annihilates $\langle \Gamma,f|$
for the particular
spin network $\Gamma$
shown in figure 1.  However, in general the result is nonzero
and the calculation 
proceeds along similar lines.  This time, one
finds only a contribution of the
form $M N \partial f (\epsilon^\alpha -\epsilon^\beta)$; that is, the
only terms that arise have both $N$ and $M$ evaluated at the same
vertex $v$.  The most
general calculation for the Rovelli--Smolin volume generates only
the terms in (\ref{smallsep}).

In all cases one sees that the Hamiltonian is lacking the ingredients
to produce the correct commutator, even if one could somehow reregulate
the commutator to produce a finite result.
In particular, one notices
the first argument of $f$ in (\ref{smallsep}) is always evaluated at
the original vertex $v$.  Thus, a term of the form ${{\partial f}
\over {\partial v}}$, in which
$f$ is differentiated with respect to its {\it first} argument, never
appears. It is clear that 
Hamiltonians of this general form which fail to move the vertex at which they
originally act cannot reproduce the classical commutator, since one is
missing the term involving $\partial_v f$.  Thus, we see that the 
proposal of \cite{Lee}, which does not involve explicitly shrinking loops
to points, also cannot reproduce this structure.

As we see from the previous calculation, the Hamiltonian's action on
the proposed space of linear functionals
is tantamount to ``eliminating a line''
between two trivalent vertices
in which, at each vertex, two of the incident
edges join to make a single smooth curve.
This in particular implies that all spin
networks without such pairs of vertices are annihilated by the
action of a single Hamiltonian and therefore by the commutator of two
Hamiltonians. To have consistency, a quantum version of ${\cal O}(N,M)$
would also have to
annihilate such states.  The point is again that we expect $V_a$ to
act as a Lie derivative, so that the only real freedom is the
regularization of the operator corresponding to $q^{ab}$.
Thus, it would appear that $\hat{q}^{ab}$ would have to annihilate
such states as well. 
One might think that Hamiltonians which both add and remove
edges would be  better
candidates to have a consistent algebra.
However, we note that the `symmetrized' Hamiltonians discussed in \cite{LeMa}
do just this, but
still do not yield an appropriate algebra \cite{LeMa}.  In particular, 
one may repeat the calculation (\ref{comm}) for the symmetrized
operators and 
find similar results.

\section{Conclusions}

In this paper we have studied two issues concerning the 
quantum version of the Hamiltonian (Wheeler--DeWitt)
constraint of 
gravity proposed by Thiemann. We have shown
that, with the regulation scheme of \cite{QSD3},
the commutator algebra of two constraints is consistent on the
space of non-diffeomorphism invariant functions introduced by
Lewandowski and Marolf, but at the price of representing
the contravariant metric $q^{ab}$ (contracted with
the diffeomorphism generator)  
by the zero operator on a rather generic class of
states which includes the familiar diffeomorphism invariant linear 
functionals.
We found a way to regularize the  object ${\cal O}(N,M)$ (\ref{O}) (the
classical right hand side of the Poisson bracket (\ref{algebra}))
that avoids this problem, but then
the commutator is anomalous. We then studied  a rescaled version of the
regulated commutator
on non-diffeomorphism invariant states and found that, in this context, 
an algebra of the classical form will not be obtained unless
the constraints
somehow move the vertices on which they act.

\section*{Acknowledgements}

We wish to thank Abhay Ashtekar for several important discussions that
crystallized the ideas in this paper.  We also wish to thank Carlo
Rovelli for correspondence and especially Thomas Thiemann for pointing
out an error in a previous version of this work and discussions.  Much
of this work was done at the workshops on quantum gravity at the
Banach Center in Warsaw and at the Erwin Schr\"odinger Institute in
Vienna. We wish to thank the respective organizers, Jerzy Kijowski,
Peter Aichelburg and Abhay Ashtekar for the opportunity to participate
in these sessions.  This work was also supported in part by grants
NSF-INT-9406269, NSF-PHY-9423950, NSF-PHY-3535722 by funds of the
Pennsylvania State University and its Office for Minority Faculty
Development, funds of the Eberly Family Research Fund at Penn State,
funds from the Max Planck Society, and by funds from Syracuse
University.  We also acknowledge the support of CONICYT and PEDECIBA
(Uruguay).  JP also acknowledges support from the Alfred P.  Sloan
Foundation through an Alfred P.  Sloan fellowship. RG acknowledges the
Associate Membership Programme of the International Centre for
Theoretical Physics at Trieste for partial support. JL thanks
Alexander von Humbolt-Stiftung (AvH), the Polish Committee on
Scientific Research (KBN, grant no.  2 P03B 017 12) and the Foundation
for Polish-German cooperation with funds provided by the Federal
Republic of Germany for the support.  DM would also like to think the
Albert Einstein Institute (Max Planck Instit\"ut f\"ur
Gravitationsphysik) for its hospitality during the writing of this
paper.


\begin{references}
\bibitem{QSD1} T. Thiemann,  ``Quantum Spin Dynamics (QSD)'' gr-qc/9606089.
\bibitem{QSD2} T. Thiemann,  ``Quantum Spin Dynamics (QSD) II'' gr-qc/9606090.
\bibitem{QSD3} T. Thiemann, ``QSD 3: Quantum constraint algebra and physical
scalar product in quantum general relativity'' gr-qc/9705017.
\bibitem{QSD} T. Thiemann, ``Quantum Spin Dynamics (QSD) 4-6'', 
preprints gr-qc/9705018-20.
\bibitem{DPR} R. Borissov, R. Di Pietri, and C. Rovelli, ``Matrix
elements of Thiemann's Hamiltonian constraint in loop quantum
gravity'' gr-qc/9703090, to appear in Class. Quan. Grav.
\bibitem{Lee} L. Smolin, ``The classical limit and the form of the
Hamiltonian constraint in nonperturbative quantum gravity''
gr-qc/9609034.
\bibitem{LeMa} J. Lewandowski, D. Marolf, in preparation.
\bibitem{RoSm} C. Rovelli, L. Smolin, Phys. Rev.. Lett. {\bf 72}, 446 (1994).
\bibitem{DePico} R. De Pietri, private communication.
\bibitem{QSD5} T. Thiemann, ``QSD V: Quantum gravity as a natural regulator
of matter quantum field theories'', gr-qc/970019.
\bibitem{ReRo} M. Reisenberger, C. Rovelli, ``Sum over surfaces form
of loop quantum gravity'' gr-qc/9612035.
\bibitem{RoSmvo} C. Rovelli, L. Smolin, Nucl. Phys. {\bf B442}, 593 (1995).
\bibitem{GaPu} J. Griego, R. Gambini, and J. Pullin,
``Chern--Simons states in spin
network quantum gravity'' Phys. Lett. {\bf B} (to appear).
\bibitem{AA} A. Ashtekar, {\it Non-Perturbative Canonical Gravity},
Lectures notes prepared in collaboration with R.S. Tate (World Scientific,
Singapore, 1991); in {\it Gravitation and Quantization}, B. Julia (ed)
(Elsevier, Amsterdam, 1995).
\bibitem{JS} T. Jacobson and L. Smolin, {\it Nucl. Phys.}
{\bf B299} 295, 1988.
\bibitem{NP} C. Rovelli and L. Smolin, {\it Nucl. Phys.}
{\bf B331} (1990) 80.
\bibitem{Baez} J. Baez, ``Spin networks in nonperturbative quantum gravity'' 
gr-qc/9504036.
\bibitem{SN} C. Rovelli and L. Smolin, {\it Phys. Rev. } {\bf D53} (1995)
5743, gr-qc/9505006. 

\bibitem{ALM}  A. Ashtekar and J. Lewandowski in {\it  Quantum Gravity and
Knots}, ed. by J. Baez, Oxford Univ. Press. 1994,
e-Print Archive: gr-qc/9311010. 

\bibitem{ALMMT}   A. Ashtekar, J. Lewandowski, D. Marolf, J. Mour\~ao, and T.
Thiemann, J. Math. Phys. {\bf 36} (1995) 6456-6493, gr-qc/9504018.


\bibitem{lat} See H. Fort, R. Gambini, J. Pullin, ``Lattice knot
theory and quantum gravity in the loop representation'', 
{\it Phys. Rev. D} {\bf 56}, 2127 (1997). 

\end{references}
\end{document}